\documentclass[12pt]{article}

\usepackage{epsf}
\usepackage{rotate}
\usepackage{cite}

\def\eps{\varepsilon}

\def\as{\alpha_s}

\newcommand{\mt}{m_{\rm t}}
\newcommand{\mtb}{\overline{m}_{\rm t}}

\newcommand{\mc}{m_{\rm c}}
\newcommand{\ms}{m_{\rm s}}

\newcommand{\mb}{m_{\rm b}}
\newcommand{\mw}{M_{\rm W}}

\newcommand{\bea}{\begin{eqnarray}}
\newcommand{\eea}{\end{eqnarray}}
\newcommand{\bd}{\begin{displaymath}}
\newcommand{\ed}{\end{displaymath}}

\newcommand{\beq}{\begin{equation}}
\newcommand{\eeq}{\end{equation}}
\newcommand{\be}{\begin{equation}}
\newcommand{\ee}{\end{equation}}

\newcommand{\order}{{\cal O}}
\newcommand{\f}{\frac}

\newcommand{\msbar}{{\overline{\rm MS}}}
\newcommand{\dsp}{\displaystyle}

\begin{document}


\thispagestyle{empty}

\rightline{TUM-HEP-295/97}
\rightline{October 1997}
\vspace*{1.2truecm}
\bigskip

\centerline{\LARGE\bf  Next-To-Leading-Order Matching}
\vspace{0.3truecm}
\centerline{\LARGE\bf for the Magnetic Photon-Penguin Operator}
\vspace{0.3truecm}
\centerline{\LARGE\bf in the $B \to X_s \gamma$ Decay}
\vskip1truecm
\centerline{\large\bf Andrzej J. Buras, Axel Kwiatkowski and Nicolas Pott}
\bigskip
\centerline{\sl  Technische Universit{\"a}t M{\"u}nchen, Physik 
Department}
\centerline{\sl D-85748 Garching, Germany}
\vspace{1.5truecm}
\centerline{\bf Abstract}
The initial condition at the matching scale
$\mu_W = {\cal O}(M_W)$ for the Wilson coefficient of the
magnetic photon-penguin
operator in the decay $B\rightarrow X_s \gamma$ is calculated in the
next-\-to-\-leading-order approximation. The technical details of the
necessary two-loop calculation in the full theory are described and 
the matching with the corresponding result in the effective theory is
discussed in detail. 
Our outcome for the initial condition confirms the
final results of Adel and Yao and Greub and Hurth. We show that ---
contrary to the claims in the second of these papers
 --- the matching procedure 
can be properly performed for infrared divergent amplitudes, i.\,e.
independently of contributions from gluon bremsstrahlung.
 
\vspace*{2.0cm}

\begin{center}
{\small 
Supported by the
German Bundesministerium f{\"u}r Bildung and Forschung under contract 
06 TM 874 and DFG Project Li 519/2-2.

}
\end{center}


\newpage


\section{Introduction}

The inclusive rare $B$ meson decay $B \rightarrow X_s \gamma$ plays an
important role in present day phenomenology. In the Standard Model
it originates at the one-loop level in the so-called magnetic
photon-penguin diagrams and receives considerable QCD
corrections \cite{bertolini:87}.
 Since it is a loop-induced decay, it is naturally
suppressed and simultaneously sensitive to physics beyond
the Standard Model. However, in order to discover some new physics in $B
\rightarrow X_s \gamma$ it is essential that both the experimental
data and the theoretical prediction in the Standard Model for this
decay reach sufficient precision.  

Experimentally, the branching ratio for $B \rightarrow X_s \gamma$ is
found by the CLEO collaboration to be \cite{CLEO2}
\beq \label{resCLEO}
{\cal B}[B \rightarrow X_s \gamma] = (2.32 \pm 0.57 \pm 0.35) \times
10^{-4},
\eeq
and a very preliminary result from the ALEPH collaboration reads \cite{ALEPH}
\beq \label{resALEPH}
{\cal B}[B \rightarrow X_s \gamma] = (3.38 \pm 0.74 \pm 0.85) \times
10^{-4}.
\eeq
In (\ref{resCLEO}) and (\ref{resALEPH}) the first error is statistical
and the second is systematic. While these results put already some
constraints on the physics beyond the Standard Model, the experimental
errors have to be considerably reduced before some firm conclusions
can be reached.

On the theoretical side, during the last five years a considerable
effort has been made to calculate the important QCD effects in this
decay, including next-to-leading-order (NLO) corrections in 
renormalization
group improved perturbation theory. One of the main motivations for
this enterprise were $\pm 25\%$ renormalization scale uncertainties
\cite{ali:93,BMMP:94} in the leading order branching ratio which --- as
anticipated in \cite{BMMP:94} --- could only be reduced by extending the
calculations beyond the leading order.

Fortunately as of 1997 the complete NLO corrections to the $B
\rightarrow X_s \gamma$ decay are known. It was a joint effort of many
groups. The ${\cal O}(\alpha_s)$ corrections to the initial conditions
for the Wilson coefficients of the relevant magnetic penguin operators
at the scale $\mu_W = {\cal O}(\mw)$ have been calculated in
\cite{adel:94} and confirmed in \cite{GH97}. The next-to-leading $8
\times 8$ anomalous dimension matrix necessary for the renormalization
group evolution of the Wilson coefficients from $\mu=\mu_W={\cal O}(M_W)$
down to $\mu=\mu_b={\cal O}(\mb)$ has been calculated in
\cite{altarelli:81,buras:90, buras:92,
ciuchini:93,misiak:95,misiak:97} of which
 \cite{altarelli:81,buras:90, buras:92,ciuchini:93,misiak:95}
  are two-loop calculations
and \cite{misiak:97} is a very difficult three-loop calculation. The
one-loop matrix elements $\langle s \gamma \vert Q_i \vert b\rangle$
and the gluon bremsstrahlung contributions $\langle s \gamma g \vert Q_i
\vert b \rangle$ have been calculated in \cite{ali:91,pott:96}. Finally,
the very difficult two-loop corrections to $\langle s \gamma \vert Q_i
\vert b \rangle$ were presented in \cite{greub:96}.

In a recent letter \cite{BKP} we have analyzed the scale uncertainties
in the $B \rightarrow X_s \gamma$ decay including not only the scale
uncertainty in $\mu_b$ considered in the papers above, but also the
uncertainties in the choice of $\mu_W$ and the choice of the scale
$\mu_t$ entering the running top quark mass $\overline{m}_t(\mu_t)$.
 To this end
we have repeated the calculation of the initial condition of the by
far dominant Wilson coefficient of the 
magnetic photon-penguin operator
$Q_7$ confirming the final result in
 \cite{adel:94, GH97} and generalizing
it to include the dependences on $\mu_t$ and $\mu_W$ with $\mu_t \not=
\mu_W$. In \cite{adel:94} and \cite{GH97} $\mu_W=\mu_t$ was
used. Our numerical analysis of the complete NLO corrections gave
\beq
{\cal B}[B \rightarrow X_s \gamma] = (3.48 \pm 0.13 \mbox{(scale)} \pm
0.28 \mbox{(par)}) \times 10^{-4} = (3.48 \pm 0.31) \times 10^{-4}
\eeq
where we show separately scale and parametric uncertainties.

The new feature of this result compared to the previous NLO analyses
\cite{misiak:97,greub:96} is the smallness of the remaining scale
uncertainties by a factor of two relative to the ones quoted in these
papers. The origin of this difference, which is related to the
numerical analysis, has been discussed in detail in \cite{BKP} and
will not be repeated here. This welcome reduction of the scale
uncertainties cannot be fully appreciated at present
 because of considerable
parametric uncertainties originating dominantly
in the charm and bottom quark
masses. We believe that these parametric uncertainties will be reduced
in the future by at least a factor of two, so that a prediction for
${\cal B}[B \rightarrow X_s \gamma]$ with an uncertainty of $5-10\%$
will be available one day. With more precise measurements of ${\cal
B}[B \rightarrow X_s \gamma]$ expected from the upgraded CLEO detector
as well as from the $B$ factories at SLAC and KEK this should allow a
useful test of the Standard Model, possibly giving some hints beyond
it.

The purpose of the present paper is the presentation of the details of
our calculation of the initial condition for the Wilson coefficient
$C_7(\mu_W)$ of the magnetic photon-penguin 
 operator. This initial
condition is obtained from a matching of the full theory (involving
the $W$ boson and the top quark) with the effective theory in which
$W$ and top do not appear as dynamical degrees of freedom: they have
been integrated out.

Since our result agrees with the previous calculations done in
\cite{adel:94} and \cite{GH97} we would like to point out right
away what is new in our paper.

Our method is very similar to the one used by Greub and Hurth
\cite{GH97} and will be explained in the following
 sections. In \cite{GH97}
dimensional regularization ($D=4-2 \eps$) is used to regulate both
infrared and ultraviolet singularities.  It has
been stressed there that in order to obtain the final result for
$C_7(\mu_W)$ it is essential
\begin{enumerate}
\item to distinguish the $1/\eps$ ultraviolet singularities from
the infrared ones, which in \cite{GH97} are denoted by
$1/\eps_{\rm IR}$,
\item that the matching between the full and the effective theory is
done in $D=4$ dimensions and that it can only be performed for
infrared finite quantities. In our case  this would mean that
also the contributions from bremsstrahlung ($b\rightarrow s\gamma g$)
have to be considered. 
\end{enumerate}
We disagree on both points.
First of all it is certainly not necessary to use different $\eps$'s
for ultraviolet and infrared divergences which simplifies considerably
the calculations and avoids the appearance of dubious terms like
$\eps/\eps_{\rm IR}$ in \cite{GH97}.
After proper renormalization of ultraviolet singularities the
left-over divergences are of infrared origin only. These
sigularities will automatically cancel in the matching procedure since
the Wilson coefficients are purely short-distance quantities,
independent of the infrared structure of the theory. This implies
contrary to Greub and Hurth that the matching of the full and the
effective theory can correctly be done for infrared divergent
quantities. In particular, as we will demonstrate in this paper, it
can be performed by considering only the virtual corrections to the
 decay $b \rightarrow s \gamma$, i.\,e. without invoking the process 
$b \rightarrow s \gamma g$.
To this end, however, it is essential to perform the matching in
$D=4-2\eps$ dimensions. 
This requires that ${\cal O}(\eps)$ terms in the Wilson coefficients
have to be kept during the calculation until all infrared divergences
present in the full theory have been cancelled in the process of
matching by those present in the effective theory. The authors of
\cite{GH97}  did not keep these  ${\cal O}(\eps)$ terms which is
correct if one restricts oneself to infrared finite
quantities. However, this restriction implies that --- at least in
principle --- the matching has to be done on the level of the decay rate
(i.\,e. including bremsstrahlung), whereas in our approach the matching
can be performed in a straight forward manner between amplitudes. We
regard this as a conceptual advantage. 
Despite of this Greub and Hurth formulated the matching condition in
their work for infrared finite amplitudes which they constructed by
 subtraction of pure $1/\eps_{\rm IR}$ poles as well as $\eps/\eps_{\rm
IR}$ terms. However, they do not show explicitly that bremsstrahlung
contributions can justify this definition.

To demonstrate our points, we think it is useful and profitable for
future calculations of this sort to provide an explicit presentation
of our NLO calculation of $C_7(\mu_W)$.    

Our paper is organized as follows: 
\noindent
In Section 2 we recall briefly the theoretical framework for the
$B\rightarrow X_s \gamma$ decay. In Section 3 we outline the matching
procedure and discuss the issue of the ${\cal O}(\eps)$-terms in the
Wilson coefficients in explicit terms. In Section 4 the calculations
in the full theory are presented. Here we also review briefly the 
Heavy Mass Expansion method. In Section 5 the corresponding
calculations in the effective theory are described. The final result
for $C_7^{(1)}(\mu_W)$ is given in Section 6. Section 7  briefly summarizes
 our paper.

\section{Theoretical Framework}

Perturbative QCD effects to the theoretical prediction for the $B\rightarrow
X_s\gamma$ decay are associated with large logarithms 
$\as^n(\mu_b)\ln^m(\mu_W/\mu_b)\; (m\leq n),$ where $\mu_W$ is a scale of
the order of $\mw$ or $\mt$ and $\mu_b\simeq \mb$ is the scale
of the hadronic decay under consideration. The resummation of these
logarithms to all orders in leading ($n=m$) or next-to-leading
($n=m+1$) approximation is achieved in the framework of a low energy
theory where all heavy particles like the $W$ boson and the top quark 
are integrated out. The effective Hamiltonian relevant for
$B\rightarrow X_s\gamma$ reads
\beq \label{HamEff}
{\cal H}^{\rm eff} = - \frac{G_F}{\sqrt{2}}\:\lambda_t\: \sum_{i=1}^8
C_i(\mu)Q_i \equiv -\frac{G_F}{\sqrt{2}}\:\lambda_t\:\vec Q^T \vec C(\mu)
\eeq  
with the CKM factor $\lambda_q=V_{qb}V_{qs}^{\ast}$.
The current-current operators $Q_1,Q_2$ and the QCD
penguin operators $Q_3,\dots,Q_6$ can be found e.\,g.\ in Eq.\ (IX.2) of
\cite{BBL}. Here
 we only list the magnetic photon-penguin and the magnetic gluon-penguin
 operators
\begin{equation}\label{DipOpe}
Q_{7}  =  \frac{e}{8\pi^2} \mb \bar{s}_\alpha \sigma^{\mu\nu}
          (1+\gamma_5) b_\alpha F_{\mu\nu},\qquad            
Q_{8}     =  \frac{g_s}{8\pi^2} \mb \bar{s}_\alpha \sigma^{\mu\nu}
   (1+\gamma_5)T^a_{\alpha\beta} b_\beta G^a_{\mu\nu}
\end{equation}
where the contributions from the mass insertions on the external $s$-quark
line are neglected due to the approximation $\ms \ll \mb$.

Although matrix elements $\langle {\cal H}^{\rm eff} \rangle$ of the effective
 Hamiltonian (\ref{HamEff}) are physical and therefore
 renormalization scale invariant quantities, the
 separate $\mu$-dependence of the Wilson coefficients $C_i(\mu)$ and the
 matrix elements $\langle Q_i(\mu) \rangle$
 reflects the factorization of short-distance and
 long-distance physics. The scale dependence of the coefficient
 functions is governed by the renormalization group equation
\beq\label{RGEcoeff}
\mu \frac{d}{ d\mu}\vec{C}(\mu) = \hat{\gamma}^T(\as)\vec{C}(\mu)
\eeq
where $\hat{\gamma}(\as)$ is the $8\times 8$ anomalous dimension
matrix. The solution of (\ref{RGEcoeff}) is given by 
\beq\label{evolution}
\vec{C}(\mu_b) = \hat{U}(\mu_b,\mu_W)\vec{C}(\mu_W)
\eeq
where $\hat{U}(\mu_b,\mu_W)$ is an evolution matrix from $\mu_W$
down to $\mu_b$ and $\vec{C}(\mu_W)$ are the initial conditions
for this evolution. An explicit formula for $\hat{U}$ in terms of
$\hat{\gamma}(\as)$ is given e.\,g. in Eq.\ (III.93) of \cite{BBL}.
The intial conditions
 $\vec{C}(\mu_W)$ for the operators $Q_1,Q_2$ and $Q_3,\dots,Q_6$,
calculated in \cite{buras:90} and 
\cite{buras:92} respectively, can also be found there. The
 purpose of the present paper is the calculation of ${\cal O}(\as)$
 corrections to $C_7(\mu_W)$.

\section{Outline of the Matching Procedure}
Since the Wilson coefficients are independent of the external states
in the matrix elements, the initial conditions may be determined by
considering the decay process at the quark level $b\rightarrow s\gamma$.
They are obtained
by matching the full and the effective theory at the matching scale
$\mu=\mu_W$. Therefore the matrix
element ${\cal M}(b\rightarrow s\gamma)$ is first to be calculated in the
full theory which contains the full particle spectrum of the Standard
Model. At the NLO level this amounts to the computation of two-loop
diagrams
as depicted in Figure~1,
\begin{figure}[thb]
\centerline{
\epsfxsize=14cm
\epsfysize=21.5cm
\epsffile{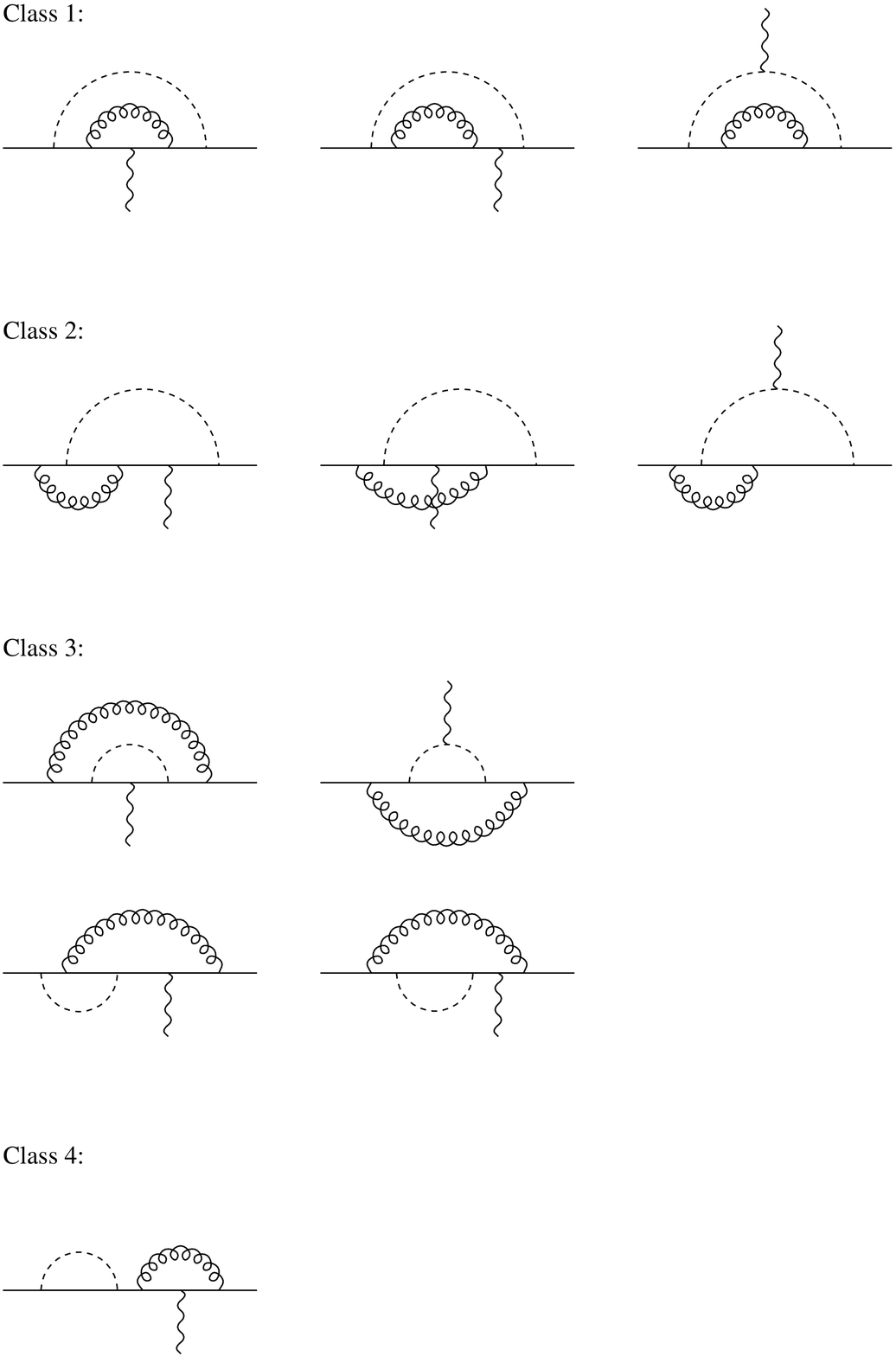}}
\caption{\small The Feynman graphs contributing to $b \rightarrow s
\gamma$ in the full theory. The internal quark is either a top or a
charm, and every dashed line can be a $W^\pm$ boson or a Higgs-ghost
$\Phi^\pm$. Left-right symmetric
diagrams and diagrams contributing only to the structure function
${\cal F}_3$ (see text) are not shown.}\label{fig1}
\end{figure} where  
 heavy particles like the top quark, $W^{\pm}$ bosons and Higgs-ghosts
$\Phi^{\pm}$ explicitly appear as virtual states.
The result has the following structure:
\beq\label{MEfull}
{\cal M}(b\rightarrow s\gamma) = -\frac{G_F}{\sqrt{2}}\:\lambda_t\:
K_7(\mu) \langle s\gamma|Q_7|b\rangle^{(0)}
\equiv
{\cal M}^{(0)}+{\cal M}^{(1)}
\eeq
where the respective terms
 ${\cal M}^{(0)}$ and ${\cal M}^{(1)}$ of 
the zeroth and first order in $\as$
 correspond to the decomposition of $K_7$
\beq\label{expand1}
K_7(\mu) = K_7^{(0)}(\mu) + \frac{\as}{4\pi}K_7^{(1)}(\mu)
.
\eeq
Furthermore,
 $\langle s\gamma|Q_7|b\rangle^{(0)}$ denotes  the tree level
matrix element of $Q_7$.

In a  second step the calculation of the matrix
element in the effective theory $\hat{\cal M}(b\rightarrow s\gamma)$
is required, where 
all heavy particles have been integrated
 out. Accordingly the 
diagrams in Figure~2
\begin{figure}[thb]
\centerline{
\epsfysize=13cm
\rotate[r]{
\epsffile{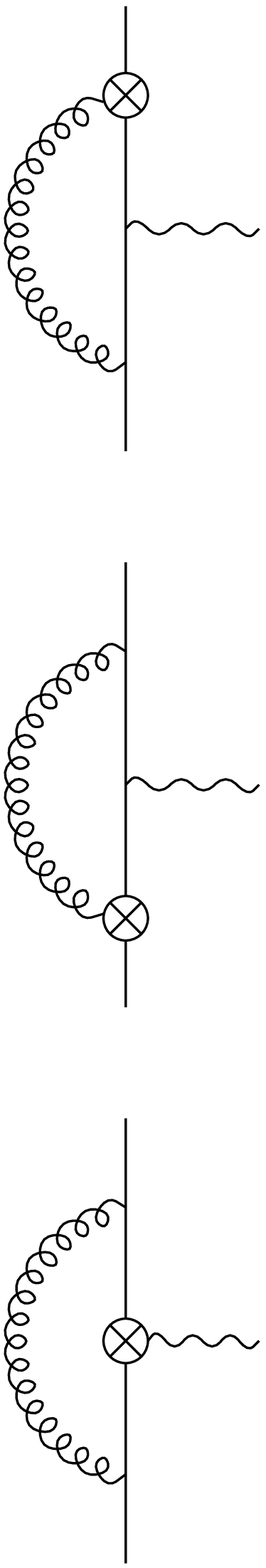}}}
\caption{\small The three $b \rightarrow s \gamma$ Feynman graphs of
the effective theory involving the operators $Q_7$ and $Q_8$. Four
additional diagrams containing $Q_8$ are not shown since they contribute only
to the structure function ${\cal F}_3$ (see text).}\label{fig2}
\end{figure} are composed  of light particles
only.  
Using  the shorthand notation $\langle Q_i \rangle \equiv    
 \langle s\gamma|Q_i|b\rangle$ and the expansions
\beq\label{expand2}
C_i(\mu) = C_i^{(0)}(\mu) + \frac{\as}{4\pi}C_i^{(1)}(\mu),
\;\;\;\;\;
\langle Q_i (\mu) \rangle = \langle Q_i  \rangle^{(0)} 
+ \frac{\as}{4\pi}\langle Q_i (\mu) \rangle^{(1)},
\eeq
the matrix element in the effective theory is written as
\beq\label{MEeff}
\begin{array}{ll}
\dsp
\hat{\cal M}(b\rightarrow s\gamma) 
&\dsp 
= \sum_{i=1}^8\hat{\cal M}_i \;\;\;
= - \frac{G_F}{\sqrt{2}}\:\lambda_t\:
\sum_{i=1}^8 C_i(\mu) \langle Q_i(\mu)\rangle
\\ &\dsp 
= -\frac{G_F}{\sqrt{2}}\:\lambda_t\:
\sum_{i=1}^8 \;\left( \:
 C_i^{(0)}(\mu) \langle Q_i\rangle^{(0)}
+
\frac{\as}{4\pi}\left[ 
C_i^{(0)}(\mu) \langle Q_i(\mu)\rangle^{(1)}
+C_i^{(1)}(\mu) \langle Q_i\rangle^{(0)}
                 \right]
    \:          \right).
\end{array}	
\eeq

The matching procedure between the full and the effective theory
establishes the initial conditions for the Wilson coefficients.
The matching scale $\mu_W$ is chosen in the regime
 $\mu_W\simeq {\cal O}(\mw,\mt)$, 
 thus giving rise only to small logarithms $\as
\ln(\mu_W/\mw)$ in the perturbative expansion. The matching condition
\beq\label{matching1}
\hat{\cal M}(\mu_W) = {\cal M}(\mu_W)
\eeq
translates into the LO and NLO identities

\begin{eqnarray}\label{matching2a}\dsp
{\rm LO:} 
&& \dsp
C_7^{(0)}(\mu_W) = K_7^{(0)}(\mu_W),
\\ \label{matching2b} \dsp
{\rm NLO:} 
&& \dsp
\left. \left. \sum_{i=1}^8 \right(
C_i^{(0)}(\mu_W) \langle Q_i(\mu_W)\rangle^{(1)}
+C_i^{(1)}(\mu_W) \langle Q_i\rangle^{(0)}
\right)
= K_7^{(1)}(\mu_W)\langle Q_7\rangle^{(0)}.
\end{eqnarray}

Anticipating the results for the various coefficients that will
explicitly be presented below, we would like to
stress once more our
 treatment of infrared singularities in the
problem. In contrast to the authors of \cite{GH97}, we have not
distinguished between infrared ($1/\eps_{\rm IR}$) and ultraviolet
($1/\eps_{\rm UV}$) poles in our calculation.
 Although our IR and UV
singularities are also dimensionally regularized, a  distinction
between $\eps_{\rm IR}$ and $\eps_{\rm UV}$ becomes
irrelevant, once $\eps$ is analytically continued from the region of
convergent integrals to abitrary values.
Nevertheless, at the end of the two-loop calculation of $K_7^{(1)}$ 
one ends up with remaning pole terms even after the UV
renormalization has been performed. These pole terms are easily
identified as IR divergences on physical grounds.
 Similar pole terms
 occur in the effective theory during the calculation of
the one-loop corrections to the operator matrix elements
 $\langle
s\gamma|Q_7|b\rangle^{(1)}$ and
 $\langle
s\gamma|Q_8|b\rangle^{(1)}$.
These IR singularities
cancel out in the matching procedure due to a compensation between the
full and the effective theory.

  From the NLO matching relation 
(\ref{matching2b}) one observes that the ${\cal O}(\eps)$ terms in
$C_i^{(0)}$  yield a finite contribution
 when being combined with the  singular piece of $\langle
s\gamma|Q_i|b\rangle^{(1)}$. 
The initial condition $C_7^{(1)}(\mu_W)$ is therefore obtained from
the NLO relation (\ref{matching2b}), but it requires the LO
matching to be performed up to $\order(\eps)$ in (\ref{matching2a})
as emphasized already in Section 1.
Decomposing the leading order Wilson coefficients accordingly,
\beq\label{LOeps}
C_i^{(0)} = C_{i0}^{(0)}+ \eps C_{i\eps}^{(0)},
\eeq 
the  results for $C_7^{(0)}$ and
 $C_8^{(0)}$ obtained from
the matching relation (\ref{matching2a}) read
\begin{equation}\label{c70}
C^{(0)}_{70} (\mu_W) = \frac{3 x_t^3-2 x_t^2}{4(x_t-1)^4}\ln x_t + 
   \frac{-8 x_t^3 - 5 x_t^2 + 7 x_t}{24(x_t-1)^3},
\end{equation}

\begin{equation}\label{c80}
C^{(0)}_{80}(\mu_W) = \frac{-3 x_t^2}{4(x_t-1)^4}\ln x_t +
   \frac{-x_t^3 + 5 x_t^2 + 2 x_t}{8(x_t-1)^3}                               
   \end{equation}
and for the
 $\eps$-terms
\beq\label{c701}
\begin{array}{ll}\dsp
C_{7\eps}^{(0)}(\mu_W)  
=
& \dsp
C_{70}^{(0)} \ln \f{\mu_W^2}{\mw^2}
+\f{7x_t(1+x_t)(5-8x_t)}{144(x_t-1)^3}
\\ \rule{0cm}{5ex}& \dsp
+\f{x_t \left(
   48-162x_t+157x_t^2-22x_t^3
        \right)}{72(x_t-1)^4} \ln x_t
+\f{x_t^2 \left(
   2-3x_t
        \right)}{8(x_t-1)^4} \ln^2 x_t,
\end{array}
\eeq
\beq\label{c801}
\begin{array}{ll}\dsp
C_{8\eps}^{(0)}(\mu_W)  
= 
& \rule{0cm}{5ex}\dsp
C_{80}^{(0)} \ln \f{\mu_W^2}{\mw^2}
-\f{x_t\left(26-93x_t+25x_t^2 \right)}{48(x_t-1)^3}
\\ \rule{0cm}{5ex}& \dsp
+\f{x_t \left(
   24-54x_t+14x_t^2-5x_t^3
        \right)}{24(x_t-1)^4} \ln x_t
+\f{3}{8}
 \f{x_t^2}{(x_t-1)^4} \ln^2 x_t
\end{array}
\eeq
respectively.
Here the notation
\beq\label{xt}
x_t=\f{\mtb^2(\mu_t^2)}{\mw^2}
\eeq
is used where $\mu_t$ is the scale chosen to define the running
top quark mass. 

The results (\ref{c70})--(\ref{c801}) are obtained by calculating the
usual one-loop magnetic penguin dia\-grams including ${\cal O}(\eps)$
terms. The results in (\ref{c70}) and (\ref{c80}) are the well known
leading order initial conditions for the operators $Q_7$ and $Q_8$
respectively. The results given in (\ref{c701}) and (\ref{c801}) agree
with the corresponding expressions in \cite{GH97} where they have been
denoted by $K_{701}$ and $K_{801}$ respectively. The authors of
\cite{GH97} kept these ${\cal O}(\eps)$ terms only in calculating the
renormalization counterterms in the full theory. As we stressed in
Section 1, in our approach
 they have to be kept in the effective theory as well.

\section{Calculation of $b\rightarrow s\gamma$ in the Full Theory}
\subsection{Preliminary Remarks}

For the calculation of the matrix element ${\cal M}(b\rightarrow
s\gamma)$ in the full theory the two-loop diagrams of Figure~1 have to
be considered. They are grouped into four different classes according
to a corresponding classification of integrals. Topologically the
diagrams come in two copies of the same  set, because the  virtually
exchanged quark may be the top or the charm quark. We neglect the
third possibility of an internal up quark
due to the smallness of
$\lambda_u=V_{ub}V_{us}^{\ast}\approx 0$. As a consequence the
unitarity relation $\lambda_t+\lambda_c+\lambda_u =0$ becomes
\beq\label{unitarity}
\lambda_c=-\lambda_t
.
\eeq

This justifies the overall CKM factor $\lambda_t$ in (\ref{HamEff}).
Working in the 't Hooft-Feynman gauge the virtual
bosons exchanged in the process are the $W^{\pm}$ bosons and the
Higgs-ghosts $\Phi^{\pm}$. The Yukawa coupling of the latter leads to
an $\mt^2/\mw^2$ enhancement of the top quark contributions. Similarly
a mass suppression follows from the Yukawa coupling to the charm quark
which we consider as massless, $\mc^2/\mw^2\approx 0$.

For the regularisation of ultraviolet and infrared divergences we use
dimensional regularisation with $D=4-2\eps$ dimensions and work with
the definition of an anticommuting $\gamma_5$ matrix
$\{\gamma_{\mu},\gamma_5\}=0$ (NDR scheme).
Furthermore, we employ the $\msbar$ renormalisation
scheme, i.\,e. we use the renormalization scale $\mu^2 e^{\gamma_E} / 4
\pi$ instead of $\mu^2$ and subtract the poles in $\eps$.

Since the construction of the eight-dimensional operator  basis in
(\ref{HamEff}) is based on the application of the equations of
motions for the operators \cite{grinstein:90},
only on-shell matrix elements are    
reproduced correctly \cite{PolSim}
by the Hamiltonian (\ref{HamEff}). In the
calculation of the diagrams we utilize the on-shell conditions
$q^2=(p_1-p_2)^2=0,\; p_i^2=m_i^2$ and apply the Dirac equation whenever
possible. Here $q^{\mu},p_1^{\mu},p_2^{\mu}$ denote the momenta of the
photon and the bottom and strange quarks respectively. The strange
mass is neglected throughout ($\ms=0$), except where it is needed for the
regularisation of mass singularities.

A complete calculation of the vertex $\Gamma^{\mu}_{bs\gamma}$ in the
full theory would
take into account all possible 128 Feynman diagrams and the full
dependence on all particle masses involved in these diagrams. This is
more information than is actually needed for the extraction of the
Wilson coefficient $C_7(\mu_W)$. First of all, in the effective theory
(\ref{HamEff}) terms of ${\cal O}(1/M^4)$ ($M=\mw,\mt$) are neglected by
definition. For the matching procedure, the vertex
$\Gamma^{\mu}_{bs\gamma}$ is correspondingly needed only up to
${\cal O}(1/M^2)$ , i.\,e.\ up to the first nontrivial order of an
expansion in the inverse heavy masses. Writing the most general
structure of the vertex $\Gamma^{\mu}_{bs\gamma}$ in terms of some
dimensionless structure functions ${\cal F}_i$ 
\beq\label{VTXstr}
i\Gamma^{\mu}_{bs\gamma} =  \f{G_F}{\sqrt{2}} \lambda_t {e \over 4
\pi^2} \, \bar{s} (p_2) (1+\gamma_5) \left( {\cal F}_1p_1^{\mu}\mb +{\cal
F}_2p_2^{\mu}\mb +{\cal F}_3\gamma^{\mu}\mb^2 \right) b(p_1), \eeq
we then see that these structure functions are needed only at zeroth order
in $\mb/\mw$, a fact that considerably simplifies the calculation.
Secondly, even some information contained in (\ref{VTXstr}) is
redundant: from gauge invariance it follows that $q_{\mu}
\Gamma^\mu_{bs\gamma} =0$, implying
\beq \label{f3}
{\cal F}_3 = -{{\cal F}_1+{\cal F}_2 \over 2}.
\eeq
Therefore the knowledge of ${\cal F}_1$ and ${\cal F}_2$ determines
the vertex completely: all 40 diagrams contributing only
 to ${\cal F}_3$ can be
neglected and are not listed in Figure \ref{fig1}. 
Using (\ref{f3}), one may finally write (\ref{VTXstr}) as
\beq \label{VTXop}
i\Gamma^\mu_{bs\gamma}=-i {G_F \over \sqrt{2}} \lambda_t {e \over 4
\pi^2}
K_7 \, \bar{s}(p_2) q_\nu \sigma^{\nu \mu} \mb (1+\gamma_5) b(p_1) = 
- {G_F \over \sqrt{2}} \lambda_t  K_7 \langle Q_7 \rangle^{(0)}
\eeq
with
\beq
K_7= -{\cal F}_3 = {{\cal F}_1+{\cal F}_2 \over 2}.
\eeq
This is the structure we already  anticipated in
(\ref{MEfull})\footnote{In deriving (\ref{VTXop}) we neglected the terms
in (\ref{VTXstr}) proportional to $q^{\mu}$ since they vanish after
contraction with the polarization vector $\epsilon_\mu$ ($\epsilon q=0$).}.

Finally, it should be stressed that our calculation was heavily
 supported by algebraic mani\-pu\-lation programs. We proceeded along
 two independent tracks, one based on FORM
 \cite{FORM} and the other
utilizing
 MATHEMATICA in combination with  TRACER \cite{TRACER}.

\subsection{Heavy Mass Expansion}
For calculating the vertex $\Gamma^{\mu}_{bs\gamma}$ up to ${\cal
O}(\mb^2/M^2)$ we employed the so-called Heavy Mass Expansion
which has developed into a widely
used industry by now \cite{HME1,HME2,HME3,HME4} (for a pedagogical
description see also \cite{HME5}). It approximates a Feynman integral
$\langle\Gamma\rangle$ by the asymptotic expansion
\beq\label{HMExp}
\langle\Gamma\rangle\;\;
 \stackrel{M \rightarrow \infty}{=} \;\;
\sum_{\gamma}
C_{\gamma}^{(M)}\; \star\; 
\langle\Gamma/\gamma\rangle
.
\eeq 
The notation of (\ref{HMExp}) is understood in the sense of the
following prescription:
\begin{itemize}
\item
Identify all {\it hard subgraphs} $\gamma$ of the diagram 
$\langle\Gamma\rangle$. A hard subgraph is defined as a 1PI part of
the diagram that contains all lines with heavy masses $\mw,\mt$. The
largest subgraph is the original diagram itself.
\item
Perform a formal Taylor series $C_{\gamma}^{(M)}$
in the integrand of these subgraphs with respect
 to its small quantities. Small
 quantities are small masses and
 the external momenta entering or leaving the subgraph.
The Taylor expansion in particular includes the small bottom mass
(the charm quark is considered as massless anyway), which cannot be
neglected due to its explicit appearance in the operator $Q_7$. 
\item
Shrink the subgraph to an effective blob in the diagram
 and insert for it the Taylor
expansion obtained in the previous step. Perform the integrations
 without any further expansions.
\item
Sum the contributions of all subgraphs.
\end{itemize}

The method is visualized for an example in Figure~3. 
\begin{figure}[ht]
\begin{displaymath}
\begin{array}{ccc}
\epsfysize=4.4cm
\rotate[r]{
\epsffile{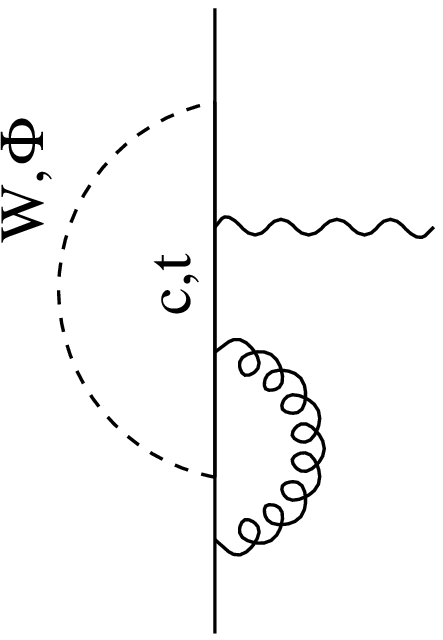}} &\hspace*{0.5cm} 
\stackrel{\displaystyle \longrightarrow}{\rule{0cm}{1.4cm}}
\hspace*{0.5cm}  &
\epsfysize=9cm
\rotate[r]{
\epsffile{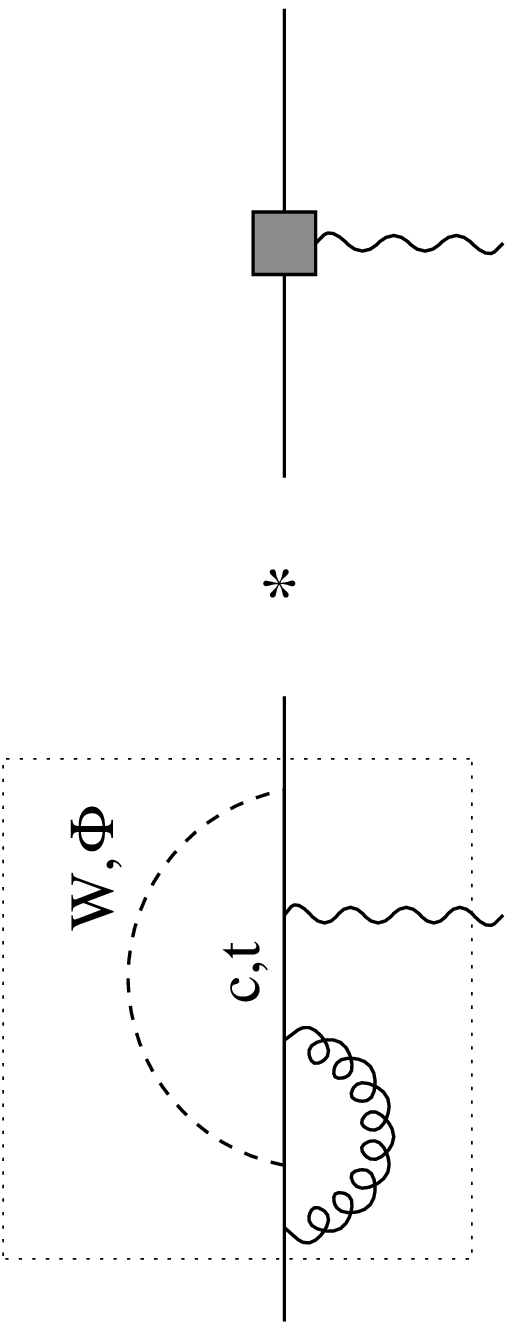}} \\
& & \hspace*{0.5cm}  C_{\gamma_1}^{(M)} \hspace*{4cm} \langle \Gamma / \gamma_1 \rangle \\
& \hspace*{1cm}\stackrel{\displaystyle +}{\rule{0cm}{1.4cm}} &
\epsfysize=9cm
\rotate[r]{
\epsffile{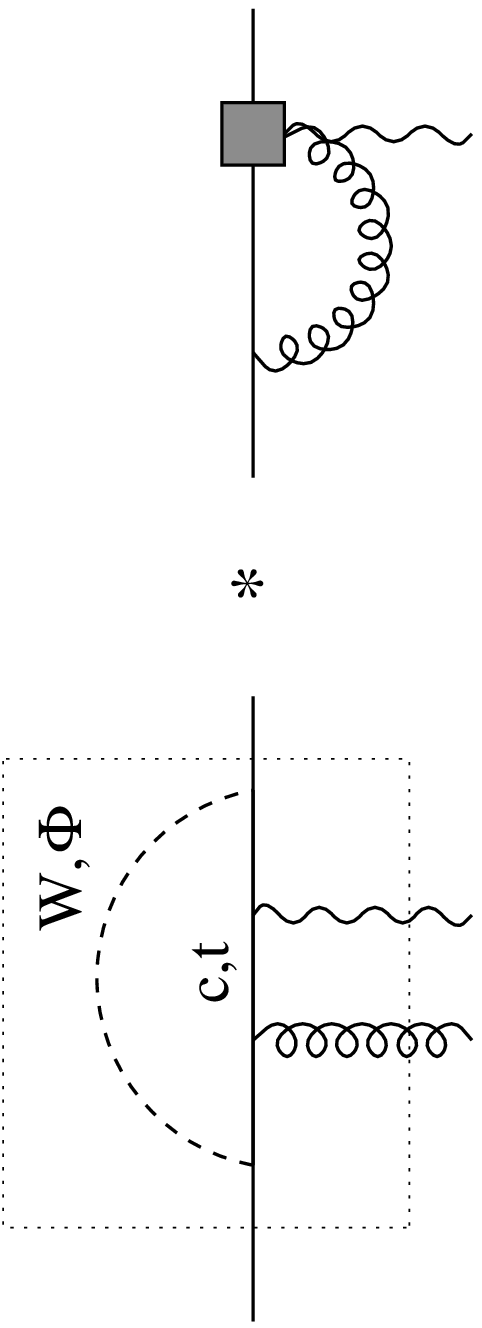}} \\
& & \hspace*{0.5cm} C_{\gamma_2}^{(M)} \hspace*{4cm} \langle \Gamma /
\gamma_2 \rangle \\
& & \\
& \hspace*{1cm} \stackrel{\displaystyle +}{\rule{0cm}{2.0cm}} &
\epsfysize=9cm
\rotate[r]{
\epsffile{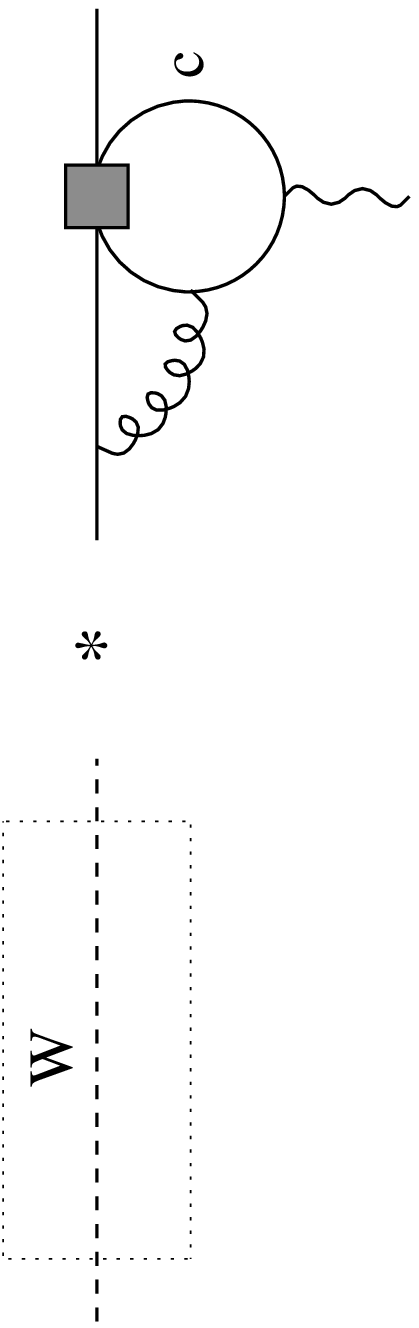}} \\
& & \hspace*{0.5cm} C_{\gamma_W}^{(M)} \hspace*{4cm} \langle \Gamma /
\gamma_W \rangle \\
\end{array}
\end{displaymath}
\caption{\small Pictorial representation of the Heavy Mass
Expansion. The third term only exists if the internal quark is a
charm.}
\end{figure}
The boxes indicate the Taylor expanded subgraphs $C_{\gamma}^{(M)}$
that are inserted in the blobs of the reduced diagrams 
 $\langle\Gamma/\gamma\rangle$.
Of the charm diagrams only those are non-vanishing which contain at
least one
$W$ boson. This is due to the Yukawa suppression $\mc^2/\mw^2=0$ of
the diagrams with Higgs-ghost exchange.

The Heavy Mass Expansion allows for an additional subdiagram in the
charm contribution as compared to the top graph.
This is the case where  either the
$W$ propagator alone or the $WW\gamma$, $W\Phi\gamma$, $\Phi W\gamma$
vertices are identified as a subgraph  $\gamma_W$
 which is not possible in a diagram with a
virtual top quark, because the top mass is considered as heavy and
  has to
be part of a subdiagram by definition. It turns out that 
it is not necessary to
explicitly compute this contribution ${\cal M}_{\rm
charm}^{(\gamma_W)}$. This is for the following reason: due to the
presence of two heavy propagators, the subgraphs $WW\gamma$,
$W \Phi\gamma$, $\Phi W\gamma$ are of ${\cal O}(1/M^4)$, so they can
be neglected for the matching with the effective theory which is accurate
only to ${\cal O}(1/M^2)$. On the other hand, the subgraphs with the
$W$ propagator alone obviously yield the same contribution as 
 the matrix element $\hat{\cal M}_2$
 of the current-curent operator $Q_2$ in the
effective theory, so both contributions completely cancel each other 
 in the matching procedure.

The contribution of the biggest, leading subgraph in the Heavy Mass
 Expansion  --- namely the one being the graph
itself --- just represents the naive expansion of the
 integrand in $\mb/M$. The virtues of the Heavy Mass Expansion are
 best seen by the way how the  obvious deficits of this too simple approach
are corrected by 
 the contributions of the subleading
 subgraphs. First, the naive expansion alone results only
  in a polynomial power expansion of the parameter  $\mb/M$, because
 by construction of the Taylor series the expansion parameter is
 nullified inside the integrals. Structures like logarithms
 $\ln(\mu^2/\mb^2)$ can only be generated from subleading 
subdiagram contributions, where the bottom mass sets the integration
 scale in the  reduced graph. Second, the expansion of the integrand
 in the leading subgraph may generate artificial infrared divergences.   
Similarly the expansion on the level of the subleading subgraph creates
 artificial divergences in the ultraviolet regime. These spurious
 divergences cancel each other in the sum of all subgraph contributions.
The cancellation is only operative if ultraviolet and infrared poles
 are identified: $\eps_{\rm UV}=\eps_{\rm IR}$. 

\subsection{Classification of Integrals}
From Figure~3 one can observe how the two-loop calculation is simplified
by the 
Heavy Mass Expansion.
 The original massive two-loop integral with external momenta either
reduces  to a massive two-loop (tadpole) integral without
external momenta or factorizes into two separate 1-loop integrals.
 
The diagrams of class 1  in Figure~1 with an internal top quark
receive their  contribution only from subdiagram $\gamma_1$ of the
Heavy Mass Expansion, namely where $\gamma_1$ represents 
the graph itself. The master integral is a  two-loop tadpole integral
with one massless and two massive lines of different mass 
\cite{BijGhin}
\begin{eqnarray}\label{bij}
\lefteqn{\dsp
\mu^{4 \eps} \int\f{d^Dp}{(2\pi)^D}\int\f{d^Dk}{(2\pi)^D}
\f{1}{(\mt^2-p^2)(\mw^2-k^2)(-[p+k]^2)}
\;\; =} 
\nonumber \\ \dsp
& \dsp
\left(\f{4 \pi \mu^2}{\mw^2} e^{-\gamma_E}\right)^{2\eps}
\left(\f{i}{16\pi^2}\right)^2\mw^2\Bigg\{
& \dsp -\f{1}{\eps^2} \f{1}{2}(1+x_t)
+\f{1}{\eps}\left[
-\f{3}{2}(1+x_t) + x_t\ln x_t
\right]
\nonumber \\
& & \dsp
+(1+x_t)\left[-\f{7}{2}-\f{\pi^2}{12}
\right]
 - (1-x_t) {\rm Li}_2(1-x_t) \nonumber
\\ & & \dsp
+3x_t\ln x_t-\f{1}{2}x_t\ln^2x_t
\Bigg\}
\end{eqnarray}
where  ${\rm Li}_2(x) = -\int_0^x {\rm d}t
\f{\ln(1-t)}{t}$.
Due to the expansion, integrals occur also with higher powers in the
denominator. 
They
 are easily obtained from (\ref{bij}) by taking derivatives
with respect to the masses.
 Higher powers in the massless
propagator require a separate calculation. 
As an example we give the result for the integral with a 
quadratic massless propagator:
\begin{eqnarray}\label{power2}
\lefteqn{\dsp
\mu^{4\eps} \int\f{d^Dp}{(2\pi)^D}\int\f{d^Dk}{(2\pi)^D}
\f{1}{(\mt^2-p^2)(\mw^2-k^2)(-[p+k]^2)^2}
\;\; =} 
\nonumber \\ \dsp
& \dsp
\left(\f{4 \pi \mu^2}{\mw^2}  e^{-\gamma_E} \right)^{2\eps}
\left(\f{i}{16\pi^2}\right)^2\Bigg\{
& \dsp -\f{1}{2\eps^2} 
+\f{1}{\eps}\left[
-\f{1}{2} + \f{x_t}{x_t-1}\ln x_t
\right]
\nonumber \\
& & \dsp
-\f{1}{2}-\f{\pi^2}{12}
 +\f{x_t+1}{x_t-1} {\rm Li}_2(1-x_t) \nonumber
\\ & & \dsp
+\f{x_t}{x_t-1}\ln x_t
-\f{x_t}{2(x_t-1)}\ln^2x_t
\Bigg\}.
\end{eqnarray}
Massless lines are raised at most to a power of 4 in the expansion.
The corresponding integrals 
are computed in an analogous manner.



For all four classes of diagrams
the subgraph $\gamma_1$ contributions 
 can be
reduced by partial fractioning
 to the above integrals and to the special case of two equal
masses
\begin{eqnarray}\label{tadpole1}
\lefteqn{\dsp
\mu^{4\eps} \int\f{d^Dp}{(2\pi)^D}\int\f{d^Dk}{(2\pi)^D}
\f{1}{(M^2-p^2)^{\alpha}(M^2-k^2)^{\beta}(-[p+k]^2)^{\gamma}}
\;\; =} \nonumber
\\ & & \dsp
\left(\f{4 \pi \mu^2}{M^2}\right)^{2\eps}
\left(\f{i}{16\pi^2}\right)^2
(M^2)^{4-\alpha-\beta-\gamma} \nonumber
\\ & & \dsp
\f{\Gamma(\alpha+\beta+\gamma-D)
\Gamma(\alpha+\gamma-D/2)
\Gamma(\beta+\gamma-D/2)
\Gamma(D/2-\gamma)}
{\Gamma(\alpha)\Gamma(\beta)\Gamma(D/2)
\Gamma(\alpha+\beta+2\gamma-D)}.
\end{eqnarray}

In case of an internal charm quark only one heavy mass $\mw$ is
 present in the subgraphs $\gamma_1$. The corresponding
  tadpole integral reads  
\begin{eqnarray}\label{tadpole2}
\lefteqn{\dsp
\mu^{4\eps} \int\f{d^Dp}{(2\pi)^D}\int\f{d^Dk}{(2\pi)^D}
\f{1}{(M^2-p^2)^{\alpha}(-k^2)^{\beta}(-[p+k]^2)^{\gamma}}
\;\; =} \nonumber
\\ & & \dsp
\left(\f{4 \pi \mu^2}{M^2}\right)^{2\eps}
\left(\f{i}{16\pi^2}\right)^2
(M^2)^{4-\alpha-\beta-\gamma}
\nonumber \\ & & \dsp
\f{
\Gamma(D/2-\gamma)
\Gamma(D/2-\beta)
\Gamma(\beta+\gamma-D/2)
\Gamma(\alpha+\beta+\gamma-D)
}
{\Gamma(\alpha)\Gamma(\beta)
\Gamma(\gamma)
\Gamma(D/2)}.
\end{eqnarray}

Different types of integrals come into play during the 
 calculation of the contributions from 
 the second subgraphs $\gamma_2$.
 Whereas the subdiagrams themselves amount to
 a 
  trivial one-loop tadpole integral, the reduced graphs
$\langle \Gamma /\gamma_2\rangle$ lead to many different kinds of massive 
one-loop integrals with nonvanishing external momenta. They are
computed in the standard manner.

As far as the reduced graphs are concerned, class 2 diagrams are
 characterized by two-point functions with the quark being either the
 bottom or the strange quark.  Some of the integrals containing the
 strange quark become singular in the small $\ms$ limit.  In these
 cases the strange mass has to be kept as a regulator, but can be
 neglected otherwise.

In classes 3 and 4 three-point functions must be calculated with
 either two different or two equal masses. In the  presence of a strange
 quark again its mass  $\ms$ serves as a regulator for mass
 singularities.

\subsection{Results}
The matrix element calculated in the full theory,
\beq
{\cal M}(b\rightarrow s\gamma) = 
{\cal M}_{\rm top}(b\rightarrow s\gamma) 
+{\cal M}_{\rm charm}(b\rightarrow s\gamma),
\eeq
receives contributions from diagrams with a virtual top and a virtual
charm quark respectively.
Because the formula for the top contribution is rather lenghty
 we do not reproduce it here, but give --- as
a separate intermediate result --- the charm part of the
two-loop matrix element alone 
($r=\ms^2/\mb^2$):
\beq\label{charm}
\begin{array}{l}\dsp
{\cal M}_{\rm charm}^{(1)} = 
-\f{G_F}{\sqrt{2}}\;\lambda_t\;
\langle s\gamma| Q_7|b \rangle^{(0)}\; \f{\as}{4\pi} 
\\ \dsp
\hphantom{xxxxx}
\cdot\left(-\f{1}{972}\right)
\Bigg\{
\f{210}{\eps}
-\f{828}{\eps}\ln r \left[
1+\eps \ln\f{\mu^2}{\mb^2}+\eps \ln\f{\mu^2}{\mw^2}
\right]\\
\hphantom{xxxxx}
\dsp
-3155 -96 \pi^2 +288 i \pi
-2526\ln r + 414\ln^2 r 
-1134\ln \f{\mu^2}{\mb^2}+1554\ln \f{\mu^2}{\mw^2}
\Bigg\}
\\ \dsp
\hphantom{xxx}
 + {\cal M}_{\rm charm}^{(\gamma_W)}.
\end{array}
\eeq
As discussed before, the subgraph piece 
 $ {\cal M}_{\rm charm}^{(\gamma_W)} $ of the Heavy Mass Expansion
will cancel in the matching and needs not to be calculated explicitly.
 
The complete (unrenormalized) two-loop result is obtained after adding
the top quark contribution to (\ref{charm}). We express it with the
help
of two auxiliary functions
\begin{eqnarray}\dsp
g_t 
& \dsp
 = 
& \dsp
-6x_t\f{\partial C^{(0)}_{70}}{\partial x_t},
\\ \label{gb} \dsp
g_b 
& \dsp
=
&\dsp
\f{x_t(3x_t-4)}{4(x_t-1)^3}\ln x_t
+\f{x_t(9-7x_t)}{8(x_t-1)^2}
.
\end{eqnarray}
One  arrives at
\beq\label{MEfullres1}
{\cal M}^{(1)}(b\rightarrow s\gamma) = 
-\f{G_F}{\sqrt{2}}\;\lambda_t\;
 \f{\as}{4\pi}
\Delta^{\rm unren}_{\rm full}
 + {\cal M}_{\rm charm}^{(\gamma_W)} 
\eeq
where
\beq\label{MEfullres2}
\Delta^{\rm unren}_{\rm full}
=C_7^{(0)}(\mu_W)
\langle  Q_7 \rangle^{(1)}
+ \f{4}{3}
\langle  Q_7 \rangle^{(0)}
\left[
g_1\f{1}{\eps}\left(
\f{\mu^2}{\mw^2}
\right)^{2\eps}
+g_2 \f{1}{2}\ln\f{\mb^2}{\mw^2}
+g_3
\right]
\eeq
with
\begin{eqnarray}\dsp
g_1 & = & \dsp
-g_b-g_t+C_{70}^{(0)}+\f{23}{27},
\\ 
g_2 & = & \dsp
2\left(
g_b+3C_{70}^{(0)}-\f{4}{3}C_{80}^{(0)}
\right)
+\f{4}{9},
\end{eqnarray}
\begin{eqnarray} \nonumber
g_3 & = & \dsp
\f{x_t(4-40x_t+61x_t^2+8x_t^3)}{6(x_t-1)^4} 
{\rm Li}_2(1-x_t)
+\f{2}{3}i\pi C_{80}^{(0)}-\f{2}{9} \pi^2 C_{80}^{(0)}
\\ \nonumber & & \dsp
+\f{x_t(-20-125x_t+149x_t^2+196x_t^3+16x_t^4)}{24(x_t-1)^5}
\ln x_t^2
\\ \nonumber & & \dsp
+\f{416-3448x_t+9431x_t^2-6273x_t^3-2826x_t^4-540x_t^5}{216(x_t-1)^5}
\ln x_t
\\ & & \dsp
+\f{-872+4352x_t-33369x_t^2+44732x_t^3+4597x_t^4}{1296(x_t-1)^4}.
\end{eqnarray}
The leading order coefficients were already given in
 (\ref{c70}), (\ref{c80}) and (\ref{c701}).
In (\ref{MEfullres2}) we have separated the term 
$C_7^{(0)}\langle  Q_7 \rangle^{(1)}$ which --- as can be seen in 
(\ref{matching2b}) --- will be cancelled precisely in the process of
 matching by the corresponding term in the effective theory. For
 completeness
the
explicit expression of $\langle  Q_7 \rangle^{(1)}$ is given in
(\ref{MEeffres7}). 
Inserting it into (\ref{MEfullres1}) one recovers the corresponding
 formula 
in \cite{GH97}.

\subsection{Renormalization}

We will now give the counterterms necessary to renormalize the full
theory in the $\overline{MS}$ scheme.

\subsubsection{Top Quark Mass Renormalization}

The simplest method to find the counterterm related to the top quark
mass renormalization is to take the result for $C_7^{(0)}$ in
(\ref{LOeps}) (including the $\eps$ terms) and replace there $\mt$
as follows
\beq
\mt \rightarrow \mt + (Z_{\mt}-1) \mt
\eeq
where
\beq
Z_{\mt} = 1- {\alpha_s \over 4 \pi} {4 \over \eps}.
\eeq
Expanding then in $\alpha_s$, extracting the coefficient of $\as/(4\pi)$
and multiplying it by $\langle Q_7 \rangle ^{(0)}$ 
 gives the counterterm to be added
to $\Delta^{\rm unren}_{\rm full}$ in (\ref{MEfullres1}):
\beq \label{mtcounter}
\Delta_{\mt}^c=   \frac{4}{3} \left [ {1 \over
\eps} g_t -6
x_t {\partial C_{7\eps}^{(0)}(\mu_W) \over \partial x_t} \right ]
\langle Q_7 \rangle ^{(0)}
.
\eeq
Another method is to perform the top mass renormalization diagram by
diagram by considering separately the renormalization of $\mt$ in the
top quark propagator as well as in its Yukawa coupling to the
Higgs-ghosts $\Phi^{\pm}$. This much more involved method has been
used in the two-loop calculations of QCD corrections to the
$Z^0$-penguin in \cite{buras:sdZ}. We have checked that this method
also gives the result in (\ref{mtcounter}). $\Delta_{\mt}^c$ in
(\ref{mtcounter}) reproduces the corresponding counterterm in
\cite{GH97}.

\subsubsection{Bottom Quark Mass Renormalization}
The bottom masses entering the amplitude are of twofold origin. 
Those $b$ quark masses that are introduced via the Dirac-equation or the
on-shell condition $p_1^2=\mb^2$ represent  the
renormalized on-shell quark mass and are not subject to
 the mass renormalization.  
 Consequently, the $\mb$ renormalization refers  only to
$\mb$ entering the $b$ quark Yukawa couplings to the Higgs-ghosts
$\Phi^\pm$. Since we work with
 on-shell $b$ quarks in our calculations we must
use the on-shell mass renormalization function
\beq
Z_{\mb}=1-{\alpha_s \over 4 \pi}  \left ( {4 \over \eps} +4 \ln
{\mu^2 \over \mb^2} +\f{16}{3} \right ).
\eeq
We therefore replace $\mb$ in the Yukawa couplings in the one-loop diagrams
contributing to $C_7^{(0)}(\mu_W)$
 as
follows
\beq \label{mbrepl}
\mb \rightarrow \mb + (Z_{\mb}-1) \mb
.
\eeq
The coefficient of $\as/(4\pi)$ then gives
 the $\mb$ counterterm to be added to
$\Delta^{\rm unren}_{\rm full}$ in (\ref{MEfullres1}):
\beq \label{mbcounter}
\Delta_{\mb}^c= \f{4}{3}
 \left [ g_b(x_t) \left ( {2
\over \eps}+2 \ln {\mu^2 \over \mb^2} + {8 \over 3} \right )
+\bar g_b(x_t) \right ] \left ( {\mu^2 \over \mw^2} \right )^\eps
\langle Q_7 \rangle ^{(0)}
.
\eeq
Here $g_b(x_t)$ is defined in (\ref{gb}) and $\bar g_b(x_t)$ is given
by
\beq
\bar g_b(x_t)={4 x_t-3 x_t^2 \over 8 (x_t-1)^3} \ln^2 x_t
+{10 x_t^3-13 x_t^2 \over 8 (x_t-1)^3} \ln x_t
+{25 x_t-19 x_t^2 \over 16 (x_t-1)^2}.
\eeq
$\Delta_{\mb}^c$ in (\ref{mbcounter}) agrees with the corresponding
result in \cite{GH97}.

\subsubsection{External Leg Wave Function Renormalization}

This renormalization can be avoided as it has to cancel with the
corresponding renormalization in the effective theory: 
this feature is not respected in the calculation of \cite{GH97} as
$\delta R_Z$ in their equation (3.13) does not equal to $\delta \hat
R_Z$ in (4.15) in the effective theory. By reintroducing ${\cal
O}(\eps)$ terms in $C_7^{(0)}$ in $\delta \hat R_Z$ one recovers
$\delta R_Z = \delta \hat R_Z$ as it should be. 

\subsection{Summary of the Result in the Full Theory}
Dropping ${\cal M}_{\rm charm}^{(\gamma_W)}$ and the contribution
of the wave function renormalization of the external quarks that will
be cancelled by  corresponding terms in the effective theory we find
the right hand side of the matching relation (\ref{matching2b}):
\beq\label{summaryfull}
K_7^{(1)}\langle Q_7\rangle^{(0)} = 
\Delta^{\rm unren}_{\rm full} + \Delta^c_{\mt} + \Delta^c_{\mb},
\eeq
where 
$\Delta^{\rm unren}_{\rm full}, \Delta^c_{\mt}$ and $ \Delta^c_{\mb}$
are given in 
(\ref{MEfullres2}), (\ref{mtcounter}), and (\ref{mbcounter}), respectively.

\section{Calculation of $b\rightarrow s\gamma$ in the Effective Theory}
\subsection{Unrenormalized Contributions }
In order to perform the NLO matching and
extract $C_7^{(1)}$  from the matching
relation (\ref{matching2b}), the knowledge of all other quantities in
(\ref{matching2b}) is required. 

Due to the identity of the 
matrix element $\hat{\cal M}_2$ of the operator $Q_2$ in the effective
theory with the
charm contribution ${\cal M}_{\rm charm}^{(\gamma_W)}$
in the full theory,
\beq
 \hat{\cal M}_2(b\rightarrow s\gamma) =  
{\cal M}_{\rm charm}^{(\gamma_W)}(b\rightarrow s\gamma),
\eeq
both contributions drop out from  the left and the right
hand side of (\ref{matching2b}).
In what follows we will discuss the remaining contributions in the
effective theory.

Since the Wilson coefficients for the operators $Q_1,Q_3\dots,Q_6$
 start at order $\as^1$
only their LO $(\as^0)$ one-loop 
  matrix elements $\langle s\gamma|Q_i|b\rangle^{(0)}$
are of relevance. Additionally 
  it was shown in
\cite{CFMRS:93} that only the one-loop
 matrix elements of $Q_5$ and  $Q_6$
do not vanish in the NDR scheme. Their contribution to 
 $\hat{\cal M}(b\rightarrow s\gamma)$ is
\beq\label{M5M6}
\hat{\cal M}_{5+6}  = 
-\f{G_F}{\sqrt{2}}\; \lambda_t\;
\left[
C_5(\mu_W)
\langle Q_5 \rangle^{(0)}
+C_6(\mu_W)
\langle Q_6 \rangle^{(0)}
\right]
\eeq
where
\beq\label{Q5Q6}
\langle Q_5 \rangle^{(0)}
=-\f{1}{3}\langle Q_7 \rangle^{(0)}
\left( \f{\mu^2}{\mb^2}
\right)^\eps,
\;\;\;\;\;\;\;
\langle Q_6 \rangle^{(0)}
=-\langle Q_7 \rangle^{(0)}
\left( \f{\mu^2}{\mb^2}
\right)^\eps ,
\eeq
and the coefficient functions $C_5,C_6$
are given by \cite{buras:92}
\beq\label{C5C6}
C_5(\mu) = \f{\as(\mu)}{4\pi} \left[
-\f{1}{9}\ln\f{\mu^2}{\mw^2}-\f{1}{6}\tilde{E}(x_t)
\right],
\;\;\;\;\;\;
C_6(\mu) = \f{\as(\mu)}{4\pi} \left[
\f{1}{3}\ln\f{\mu^2}{\mw^2}+\f{1}{2}\tilde{E}(x_t)
\right]
\eeq
with
\beq\label{inamiE}
\tilde{E}(x_t) = -\f{2}{3}\ln x_t +
\f{x_t^2(15-16 x_t+4 x_t^2)}{6(1-x_t)^4}\ln x_t
+\f{x_t(18-11 x_t-x_t^2)}{12(1-x_t)^3}-\f{2}{3}
.
\eeq
Since the matrix elements of $Q_5$ and $Q_6$ are finite, the possible
${\cal O}(\eps)$ terms in $C_5$ and $C_6$ can be set to zero.
Similarly the ${\cal O}(\eps)$ terms in (\ref{Q5Q6}) can be
 omitted. We will need them, however, in Section 5.2.2 in the process
 of renormalization in the effective theory. Removing the 
 overall factor in (\ref{M5M6}) and $\as/(4\pi)$
we obtain the contribution of the
 operators $Q_5$ and $Q_6$ to the left hand side of the matching
 condition in (\ref{matching2b}):
\beq\label{sum56}
\hat{\Delta}_{5+6} = -\f{4}{9}\left(
\f{2}{3}\ln\f{\mu^2}{\mw^2}+\tilde{E}(x_t)
\right)
\langle Q_7\rangle^{(0)}.
\eeq

The matrix element of the operator $Q_8$ is zero at tree level, thus
making the knowledge of $C_8^{(1)}$ superfluous for the matching.
 
One remains left with 
 the one-loop  matrix elements of the 
  operators  $Q_7$ and $Q_8$. They are calculated in the effective
 theory where 
the diagrams in Figure~2 have to be considered. 
The circles indicate the insertions of the operators.   
The computation of the corresponding massive three-point function
leads to the following 
results for the (unrenormalized) one-loop matrix elements
\cite{ali:91,pott:96}
\begin{eqnarray}\label{MEeffres7}\dsp
\langle Q_7\rangle^{(1)} 
& = & \dsp
\f{4}{3}\left[ 
-\f{1}{\eps}  \ln r
+\f{1}{2}\ln^2 r
- \ln r \ln \f{\mu^2}{\mb^2}
-2\ln r
\right]\langle Q_7\rangle^{(0)},
\\ \dsp\label{MEeffres8}
\langle Q_8\rangle^{(1)} 
& = & \dsp
\left(-\f{4}{27}\right)
\left[ 
-\f{12}{\eps} 
-33 + 2\pi^2
-12 \ln \f{\mu^2}{\mb^2}
-6i\pi
\right]\langle Q_7\rangle^{(0)}
.
 \end{eqnarray}
 Again the strange quark mass acts as an infrared regulator in
$r=(\ms^2/\mb^2)$.
In summary the left hand side of the matching condition in
(\ref{matching2b}) prior to the renormalization is given as follows
\beq\label{summaryeffunren}
\Delta_{\rm eff}^{\rm unren} = \hat{\Delta}_{5+6} 
+ C_7^{(0)}(\mu_W)\langle Q_7\rangle^{(1)}
+ C_8^{(0)}(\mu_W)\langle Q_8\rangle^{(1)}
+ C_7^{(1)}(\mu_W)\langle Q_7\rangle^{(0)}
\eeq
where $C_7^{(1)}(\mu_W)$ is the coefficient we are looking for.
Since the matrix elements
(\ref{MEeffres7}), (\ref{MEeffres8})
 contain divergences, it is necessary to
keep the ${\cal O}(\eps)$ terms in $C_7^{(0)}$ and $C_8^{(0)}$
in (\ref{summaryeffunren}).

\subsection{Renormalization}

\subsubsection{Bottom Quark Mass Renormalization}

In the effective theory the amplitude receives a contribution from the
term
\beq
C_7^{(0)}(\mu_W) \langle s \gamma \vert Q_7 \vert b \rangle^{(0)}
\eeq
where, as seen in (\ref{DipOpe}), $Q_7$ is linear in $\mb$. Making the
replacement (\ref{mbrepl}) in $Q_7$ one generates a counterterm
to be added to the left hand side of the matching condition 
(\ref{matching2b}):
\beq\label{mbcountereff}
\hat \Delta_{\mb}^c = - \left ( {4 \over \eps}
+ 4 \ln {\mu^2 \over \mb^2}+\f{16}{3} \right ) C_7^{(0)}(\mu_W)
\langle Q_7 \rangle^{(0)}
.
\eeq
This result agrees with the corresponding result in \cite{GH97} except
for the ${\cal O}(\eps)$ terms in $C_7^{(0)}(\mu_W)$ which have
been omitted  by these authors.

\subsubsection{Operator Renormalization}
The renormalization of the operator matrix elements involves the
operator renormalization and the wave function renormalization of the
quark fields. We do not include the latter as this renormalization
cancels in the process of matching with the corresponding
renormalization in the full theory. We have discussed this issue in
Section 4.5.3.

Since the operator renormalization is  ${\cal O}(\as)$, only the
counterterms for the matrix elements of the operators $Q_2,Q_7$ and
$Q_8$ have to be considered. These counterterms can easily be
calculated by using the leading order operator renormalization
constants \cite{CFMRS:93} or directly by using the relevant
coefficients of $\as/(4\pi)$ in the leading anomalous dimension matrix
$\hat{\gamma}^{(0)}$.

For the operators $Q_7$ and $Q_8$ the counterterms contributing to the
left hand side of the matching condition (\ref{matching2b}) are simply
given by
\beq\label{op77counter}
\hat{\Delta}^c_{77} = 
\f{1}{\eps}\f{\gamma^{(0)}_{77}}{2}C_7^{(0)}
  \langle Q_7 \rangle^{(0)},
\;\;\;\;\;\;
\hat{\Delta}^c_{87} = 
\f{1}{\eps}\f{\gamma^{(0)}_{87}}{2}C_8^{(0)}
  \langle Q_7 \rangle^{(0)}
\eeq
where
$\gamma^{(0)}_{77}=32/3$ and $\gamma^{(0)}_{87}=-32/9$.
The corresponding counterterms for $Q_2$, related
to its mixing under renormalization with the operators $Q_5$ and
$Q_6$,
are given by (with $C_2^{(0)}=1$)
\beq\label{op25counter}
\hat{\Delta}^c_{25} = 
\f{1}{\eps}\f{\gamma^{(0)}_{25}}{2}
  \langle Q_5 \rangle^{(0)},
\;\;\;\;\;\;
\hat{\Delta}^c_{26} = 
\f{1}{\eps}\f{\gamma^{(0)}_{26}}{2}
  \langle Q_6 \rangle^{(0)}
\eeq
with $\langle Q_5 \rangle^{(0)}$ and $\langle Q_6 \rangle^{(0)}$
given in (\ref{Q5Q6}) and
 $\gamma^{(0)}_{25}=-2/9,\;\gamma^{(0)}_{26}=2/3$.

The leading anomalous dimension $\gamma_{27}^{(0)}$ related to the
mixing of the operators $Q_2$ and $Q_7$ is obtained from two-loop
calculations as opposed to (\ref{op77counter}) and (\ref{op25counter})
which involve one-loop calculations. Because this time the mixing of
operators with different dimensions (in $D \not= 4$ dimensions of
space-time) is considered, the
relation between the relevant operator renormalization constant and
 $\gamma_{27}^{(0)}$ is more involved \cite{grinstein:90}. One finds:  
\beq\label{op27counter}
\hat{\Delta}^c_{27} = 
\f{1}{\eps}\f{\gamma^{(0){\rm NDR}}_{27}}{4}
  \langle Q_7 \rangle^{(0)}
\eeq
with $\gamma_{27}^{(0){\rm NDR}}=464/81$.

\subsection{Summary of the Result in the Effective Theory}
Adding the counterterms (\ref{mbcountereff}), (\ref{op77counter}), 
(\ref{op25counter}) and (\ref{op27counter}) to the unrenormalized result
in (\ref{summaryeffunren}) we obtain the final result for the left
hand side of the matching condition (\ref{matching2b})
\beq\label{summaryeff}
\sum_{i=1}^8 \left(
C_i^{(0)} \langle Q_i \rangle^{(1)}
+C_i^{(1)} \langle Q_i\rangle^{(0)}
\right)
=
\Delta_{\rm eff}^{\rm unren} + \hat{\Delta}^c_{\mb}
+\hat{\Delta}^c_{77}+\hat{\Delta}^c_{87}+\hat{\Delta}^c_{25}
+\hat{\Delta}^c_{26}+\hat{\Delta}^c_{27}.
\eeq

\section{Result for the Initial Condition $C_7(\mu_W)$}
The initial condition $C_7^{(1)}(\mu_W)$
 can now be determined from the NLO matching relation
 (\ref{matching2b}) by inserting (\ref{summaryfull}) and
 (\ref{summaryeff}) and solving for the unknown $C_7^{(1)}(\mu_W)$.  
 One
arrives at the final result
\beq\label{GENC7}
C_7^{(1)}(\mu_W)= C_7^{(1)}(\mw)+
8 x_t \f{\partial C_7^{(0)}(\mu_W)}{\partial x_t}\ln\f{\mu_t^2}{\mw^2} 
+\left(\f{16}{3}C_7^{(0)}(\mu_W)-\f{16}{9} C_8^{(0)}(\mu_W)
+\f{\gamma_{27}^{(0){\rm NDR}}}{2}\right) \ln \frac{\mu_W^2}{\mw^2}
\eeq
where 
\bea
C_7^{(1)}(\mw) &=& \f{2x_t(4 -40x_t +61 x_t^2 +  8x_t^3)}
{9 (x_t-1)^4} 
{\rm Li}_2 ( 1 - x_t)
\nonumber \\ &&
    +\f{x_t(-4 - 40 x_t + 37 x_t^2+71x_t^3+8x_t^4)}{9 (x_t-1)^5} \ln^2 x_t 
\nonumber \\ &&
                  +\f{2(116 - 742 x_t + 1697 x_t^2 - 1158 x_t^3
- 294 x_t^4 - 51x_t^5)} {81 (x_t-1)^5} \ln x_t
\nonumber \\ &&
                  +\f{-580  + 3409 x_t - 12126 x_t^2 + 12961 x_t^3 
 + 1520x_t^4}
{486 (x_t-1)^4} 
\eea
and $\gamma^{(0){\rm NDR}}_{27}=464/81$.
Thus we confirm the findings of \cite{adel:94} and \cite{GH97}. Our
formula generalizes the results of these papers
in the sense that we distinguish between
the  scale dependence of the matching scale $\mu_W$ and the dependence
on the mass scale $\mu_t$ at which the running top quark mass
$\overline{m}_{\rm t}(\mu_t)$ is defined. Both scales are  not
necessarily equal, but 
 were identified $\mu_W=\mu_t=\mu_{Wt}$  in \cite{adel:94,GH97}.
In \cite{BKP} the separate cancellation of these  
scales dependent terms 
in the sum of the LO and NLO contribution is demonstrated
  and the phenomenological implications are discussed.

For completion we also quote the results for the contribution of the
magnetic gluon-penguin 
 operator as calculated in \cite{adel:94,GH97} and generalized to
$\mu_W\neq\mu_t$
in \cite{BKP}
\beq\label{GENC8}
C_8^{(1)}(\mu_W)= C_8^{(1)}(\mw)+
8 x_t \f{\partial C_8^{(0)}(\mu_W)}{\partial x_t}\ln\f{\mu_t^2}{\mw^2} 
+\left(\f{14}{3}C_8^{(0)}(\mu_W)
+\f{\gamma_{28}^{(0){\rm NDR}}}{2}\right) \ln\frac{\mu_W^2}{\mw^2}
\eeq
with
\bea
C_8^{(1)}(\mw) &=& \f{x_t(-1-41 x_t -40 x_t^2 + 4 x_t^3)}
{6 (x_t-1)^4} 
{\rm Li}_2 ( 1 - x_t)
\nonumber \\ &&
      +\f{x_t(1 -146 x_t - 103 x_t^2-44x_t^3+4x_t^4)}{12 (x_t-1)^5} \ln^2 x_t 
\nonumber \\ &&
      +\f{304 -2114 x_t + 3007 x_t^2 +4839 x_t^3 + 1086
x_t^4 - 210 x_t^5} {216 (x_t-1)^5} \ln x_t
\nonumber \\ &&
        +\f{-652+1510 x_t -29595 x_t^2 - 13346 x_t^3 + 611 x_t^4}
{1296 (x_t-1)^4}
\eea
and $\gamma^{(0){\rm NDR}}_{28}=76/27$.

In the phenomenological applications it is more convenient to work
with the effective coefficients \cite{BMMP:94}
\beq\label{effcoeff8}
C_7^{{\rm eff}}(\mu_W)=C_7(\mu_W)-\f{1}{3}C_5(\mu_W)-C_6(\mu_W)
\eeq
and
\beq\label{effcoeff7}
C_8^{{\rm eff}}(\mu_W)=C_8(\mu_W)+C_5(\mu_W)
.
\eeq
Using (\ref{C5C6}), (\ref{GENC7}) and
(\ref{GENC8}) one finds the effective coefficients 
used in \cite{BKP}. For $\mu_t=\mu_W=\mw$ these effective coefficients
reduce to the ones given in \cite{misiak:97}.

\section{Summary}
In this paper we have calculated the initial condition for the Wilson
coefficient of the magnetic photon-penguin operator at next-\-to-\-leading
order. Our result agrees with the ones presented in
\cite{adel:94,GH97}. The method used by us is very different from
 the
one of Adel and Yao
 \cite{adel:94}, but is rather similar to the one of Greub and Hurth
\cite{GH97}.
However, in contrast to the latter authors, we are able to obtain the
final result by calculating only the virtual corrections to
$b\rightarrow s\gamma$ without invoking the gluon bremsstrahlung
$b\rightarrow s\gamma g$ as done in \cite{GH97}. To this end it was
necessary to keep in the process of matching  ${\cal O}(\eps)$
terms in the leading order Wilson coefficients. Our main result, which
generalizes the results of \cite{adel:94,GH97} to $\mu_W\neq\mu_t$, is
given in (\ref{GENC7}). The numerical analysis of these formulae in
the context of a complete NLO analysis of the decay $B\rightarrow
X_s\gamma$ has been presented in \cite{BKP}.

\bigskip

\vfill\eject

\end{document}